\documentclass[mathleft
]{an}
\usepackage{graphicx}
\usepackage{times}
\usepackage{natbib}
\bibpunct{(}{)}{;}{a}{}{,}
\usepackage{amsmath,amssymb}
\usepackage{lscape}
\usepackage{subfigure}
\usepackage{multirow}
\usepackage{url}
\usepackage{threeparttable}

\begin{document}

\Pagespan{1}{}
\Yearpublication{}%
\Yearsubmission{}%
\Month{}%
\Volume{}%
\Issue{}%

\title{Variability of the X-ray Broad Iron Spectral Features in Active Galactic Nuclei and Black-hole Binaries}

\author{Misaki Mizumoto\inst{1,2}\fnmsep\thanks{Corresponding author:
  \email{mizumoto@astro.isas.jaxa.jp}\newline}
, Ken Ebisawa\inst{1,2}, Masahiro Tsujimoto\inst{1},\and Hajime Inoue\inst{3}
}
\titlerunning{Variability of X-ray spectra in AGN and BHB}
\authorrunning{M.\ Mizumoto et al.}
\institute{
Institute of Space and Astronautical Science (ISAS), Japan Aerospace Exploration Agency (JAXA), 3-1-1 Yoshinodai, Chuo-ku, Sagamihara, Kanagawa, 252-5210, Japan
\and 
Department of Astronomy, Graduate School of Science, The University of Tokyo, 7-3-1 Hongo, Bunkyo-ku, Tokyo, 113-0033, Japan
\and 
Meisei University, 2-1-1 Hodokubo, Hino, Tokyo, 191-8506, Japan
}

\received{}
\accepted{}
\publonline{later}

\keywords{galaxies: active -- X-rays: binaries -- black hole physics}

\abstract{%
The ``broad iron spectral features'' are often seen in X-ray spectra of Active Galactic Nuclei (AGN) and black-hole binaries (BHB).
These features may be explained either by the ``relativistic disc reflection'' scenario or the ``partial covering'' scenario:
It is hardly possible to determine which model is valid from time-averaged spectral analysis.
Thus, X-ray spectral variability has been investigated to constrain spectral models.
To that end, it is crucial to study iron structure of BHBs in detail at short time-scales,
which is, for the first time, made possible with the Parallel-sum clocking (P-sum) mode of XIS detectors on board Suzaku.
This observational mode has a time-resolution of 7.8~ms as well as a CCD energy-resolution.  
We have carried out systematic calibration of the P-sum mode, 
and investigated spectral variability of the BHB GRS 1915$+$105.
Consequently, we found that the spectral variability of GRS 1915$+$105 does {\em not} show iron features at sub-seconds.
This is totally different from variability of AGN such as 1H0707--495, where the variation amplitude significantly drops at the iron K-energy band.
This difference can be naturally explained in the framework of the ``partial covering'' scenario.
}

\maketitle

\section{Introduction}
Many black-hole (BH) objects, both active galactic nuclei (AGN) and black-hole binaries (BHB), are known to show ``broad iron spectral features'' in their X-ray spectra around $\sim7$~keV.
These spectral features have been found in the X-ray spectrum of MCG--6--30--15 \citep{tan95},
and in many other AGN and BHB with CCD detectors such as EPIC on XMM-Newton and XIS on Suzaku (\citealt{mil07} for a review).
In this way, X-ray CCD detectors have greatly contributed to quantitative spectral analysis of the features with moderate energy resolution.
Several different models have been proposed to explain the features.
On one hand, in the ``relativistic disc reflection model'', they are interpreted as due to relativistically blurred inner-disc reflection at around an extreme Kerr BH \citep{fab89,fab02b}.
This model needs a compact ``lamp-post'' corona just above the central BH and an extremely high spin parameter, thus the X-ray emitting region must be very compact ($\lesssim1\,r_s$).
On the other hand, the ``partial covering model'' may also explain the features.
In this model, X-ray absorbers, such as outflow gas, partially cover the X-ray source, and make a strong absorption edge (e.g.~\citealt{tan03}; \citealt{miy12}).
If the emitting region is compact, absorbers cannot ``partially'' cover the X-ray source.
Thus, this model needs an extended X-ray source ($\gtrsim10\,r_s$).
Both of these models can explain the same spectral feature, thus it is difficult to determine which model is valid only using time-averaged spectra.

Therefore, various time-resolved spectral analysis methods were employed to constrain spectral models.
\citet{mat03} created root-mean-square (RMS) spectra of MCG--6--30--15 at long timescales, and found that the RMS variability is significantly reduced around 6~keV at timescales of $\sim10^{4-5}$~s.
Both models can explain the RMS spectra (e.g.~\citealt{fab03,ino03}).
Here, assuming that the spectral variability is normalized by the BH masses,
we can expect that BHBs show similar characteristic spectral variability as AGN at timescales of sub-seconds.
In order to evaluate spectral variability of BHBs quantitatively, both high energy-resolution and high time-resolution of $\sim$ms are needed.
However, such observations have been so difficult because the read-out time is much longer (e.g.\,2~s in Suzaku XIS Normal mode) in the imaging mode of CCD detectors.

Here, we use Suzaku XIS Parallel-sum clocking (P-sum) mode
to achieve both good time-resolution and good energy-resolution.
In the P-sum mode, events are stacked along the Y-axis in the CCD detector
and, in the subsequent process, the stacked events are treated as a single row.
Therefore, we can get a higher time-resolution (8~s~/~1024~$\simeq$~7.8~ms) than the Normal mode
at a cost of the spatial information along the Y-direction. 
Recently, we have improved the calibration of P-sum mode significantly in cooperation with the Suzaku XIS team\footnote{see \url{http://www.astro.isas.ac.jp/suzaku/analysis/xis/psum_recipe/Psum-recipe-20150326.pdf} for details.}, therefore
we can investigate short-time variability of X-ray spectra obtained in P-sum mode.
In this paper, we compare spectral variability of AGN and BHB with the same spectro-temporal analysis method, and
show that the spectral variability is definitely different between them.

\begin{figure}[tbp]
\centering
\includegraphics[width=55mm,angle=270]{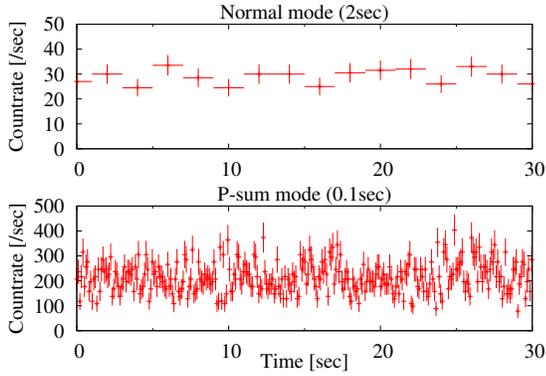}
\caption{Light curves of GRS 1915$+$105 obtained by Normal mode with 1/4 window option (Upper) and P-sum mode (Lower).}
\label{ltcrv}
\end{figure}

\section{Difference variation function method}
Most X-ray sources show temporal/spectral variability on a wide range of timescale, and various methods have been employed to investigate spectral variability.
The RMS spectrum is one of the most popular method to investigate spectral variability, but
it in principle needs continuous data that has sufficient quality \citep{ino11}.
If we try to calculate RMS spectra with small time-bins or small energy-bins,
variability would be dominated by statistical noise, and
it is hard to get information about variability of the X-ray source itself.
Therefore, it is often difficult to investigate BHB's short-time variability with this method.

Therefore, we applied the ``difference variation function method (DVF method)'', which is a newly-developed spectro-temporal data analysis method \citep{ino11}.
The DVF method we employed is as follows:
(1)~We determined the time scale $\Delta T$ and created a light curve with a bin width of $\Delta T$. 
(2)~We defined a bright/faint phase from every two adjacent bins, so that the count of the former is larger than the latter. 
(3) We created the bright/faint spectra from the bright/faint phase.
(4) Finally, we calculated the variation amplitude in the following equation:
\begin{equation}
F_\mathrm{var}=\frac{a_B-a_F}{a_B+a_F},
\end{equation}
where $a_B$ and $a_F$ show counts of the bright/faint phase, respectively.
DVF method has a great merit of being able to keep the statistics even at short timescales.
This method can extract spectral variability correlated to the observed X-ray flux.
\citet{miy12} effectively applied this method to the Suzaku data of MCG--6--30--15, and 
investigated nature of the iron K-spectral feature.
Here, we apply the DVF method to the data of both AGN and BHB, and investigate their spectral variability.

\section{Spectral variation in AGN}
\begin{figure}[tbp]
\centering
\includegraphics[width=55mm,angle=270]{1H0707.eps}
\caption{X-ray spectrum of 1H0707--495. The model line is power-law convolved with foreground absorption. The lower panel shows the ratio of the spectrum to the model.}
\label{1H0707}
%
\includegraphics[width=55mm,angle=270]{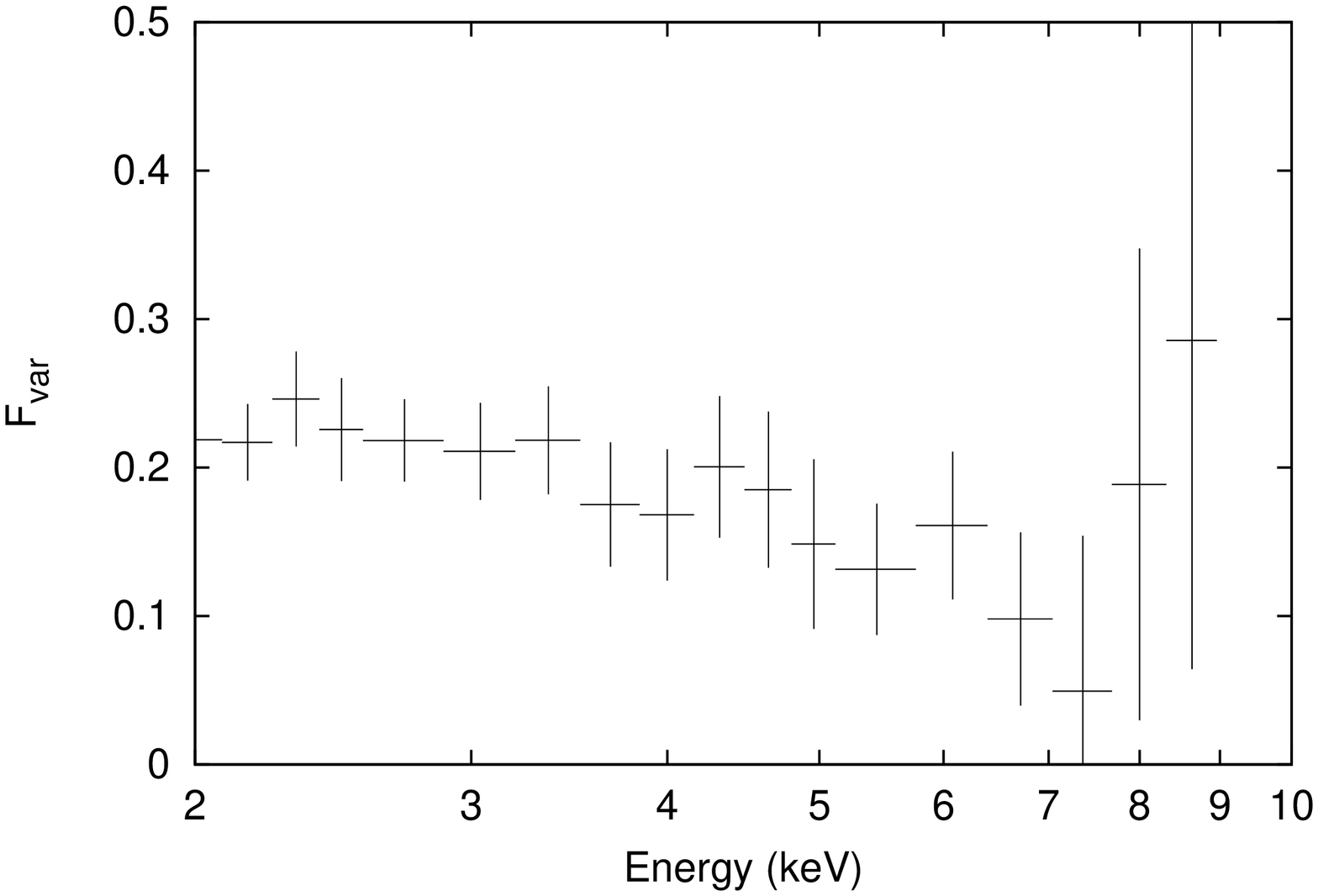}
\caption{Spectral variations of 1H0707--495. The vertical axis shows the variation amplitude with $\Delta T=8000$~s.}
\label{1H0707_DVF}
\end{figure}

As for AGN, we have analyzed the data of 1H0707--495 ($M\simeq10^6M_\odot$), which is known to have a strong ``broad iron spectral feature'' and exhibit significant X-ray time variations (see e.g.\,\citealt{miz14} and references therein).
We used the archival XMM-Newton data obtained in 2010 (ID=0653510301-601), the total exposure time of which is 399~ks.
Figure \ref{1H0707} shows the X-ray spectrum of 1H0707--495, and Figure \ref{1H0707_DVF} shows the variation amplitude at a timescale of 8000~s.
In Figure \ref{1H0707_DVF}, we can clearly see that the variation amplitude drops at $\sim7$~keV, which is consistent with \citet{gal04} and \citet{fab04}.

In the relativistic disc reflection model, this unvarying feature of the iron line is explained by the ``light bending'' model \citep{fab03,min03,min04}.
In this model, the X-ray energy spectrum is mainly composed of two components; the power law component (PLC) and the reflection-dominated component (RDC),
where the broad iron line is associated with the RDC.
Variability is mainly explained by variation the height of the ``lamp-post'' corona above the accretion disc (source height).
When the source height is small, a large part of the photons emitted from the corona are bent towards the central BH or the accretion disc with its strong gravitational field, thus PLC is weak and RDC is dominant.
As the source height gets larger, the number of photons that can escape the gravitational field increases, thus PLC gets stronger and dominant.
In this way, PLC is significantly variable as the source height varies.
On the other hand, variability of RDC is reduced by the effect of light bending.
Thus, variation amplitude around the iron K-line drops.

\begin{figure}[tbp]
\centering
\includegraphics[width=80mm]{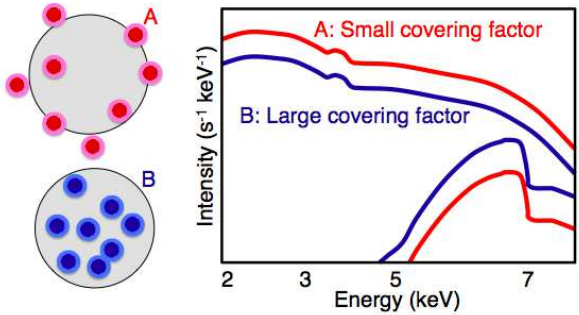}
\caption{Schematic picture of spectral variations in the partial covering model.
When the covering fraction is small (A), the direct component is bright and the absorbed component is faint. When the covering fraction is large (B), the former is faint and the latter is bright. These variations are cancelled at around the iron K-edge.
}
\label{schematic}
\end{figure}

On the other hand, in the partial covering model, the unvarying feature is explained by the iron K-edge \citep{ino03,iso15}.
Figure \ref{schematic} shows a schematic picture of spectral variations in the ``partial covering'' model.
Assuming that the variation timescale of the intrinsic luminosity is much longer than that of the partial absorbers, as the covering fraction is getting larger (A$\rightarrow$B), the direct component is getting fainter, and the absorbed component is getting brighter.
Intensities of the two components are anti-correlated; in particular, around the iron K-energy band, their variations are cancelled.
In this way, variation amplitude drops at $\sim7$~keV, and the unvarying feature is explained.
In this model, only a narrow iron emission line may exist.

\section{Spectral variation in BHB}
As for BHB, we have analyzed the data of GRS 1915$+$105 ($M\simeq10M_\odot$).\footnote{For details of data reduction and results of spectro/temporal analysis of GRS 1915$+$105, see \citet{miz15}.}
GRS 1915$+$105 shows unique and dramatic temporal/spectral variations (\citealt{fen04} for a review).
In particular, in ``State C'', which corresponds to the ``low-hard'' state of canonical BHBs \citep{bel00}, 
this object is known to have the ``broad iron spectral feature'' \citep{nei09,blu09}.
Thus, we used the archival Suzaku P-sum mode data of GRS 1915$+$105 in ``State C'' obtained on May 7--9, 2007 (ID$=$402071010).
The exposure time is 65.7~ks.
XIS1 was Normal mode (1/4~window $+$ 1~s burst option) and XIS0, 3 were P-sum mode.
The data reduction of P-sum mode followed the analysis recipe released by the XIS team$^1$.

\begin{figure}[tbp]
\centering
\includegraphics[width=55mm,angle=270]{GRS1915.eps}
\caption{X-ray spectrum of GRS1915$+$105. The model line is power-law convolved with foreground absorption. The lower panel shows the ratio of the spectrum to the model.}
\label{GRS1915}

\includegraphics[width=55mm,angle=270]{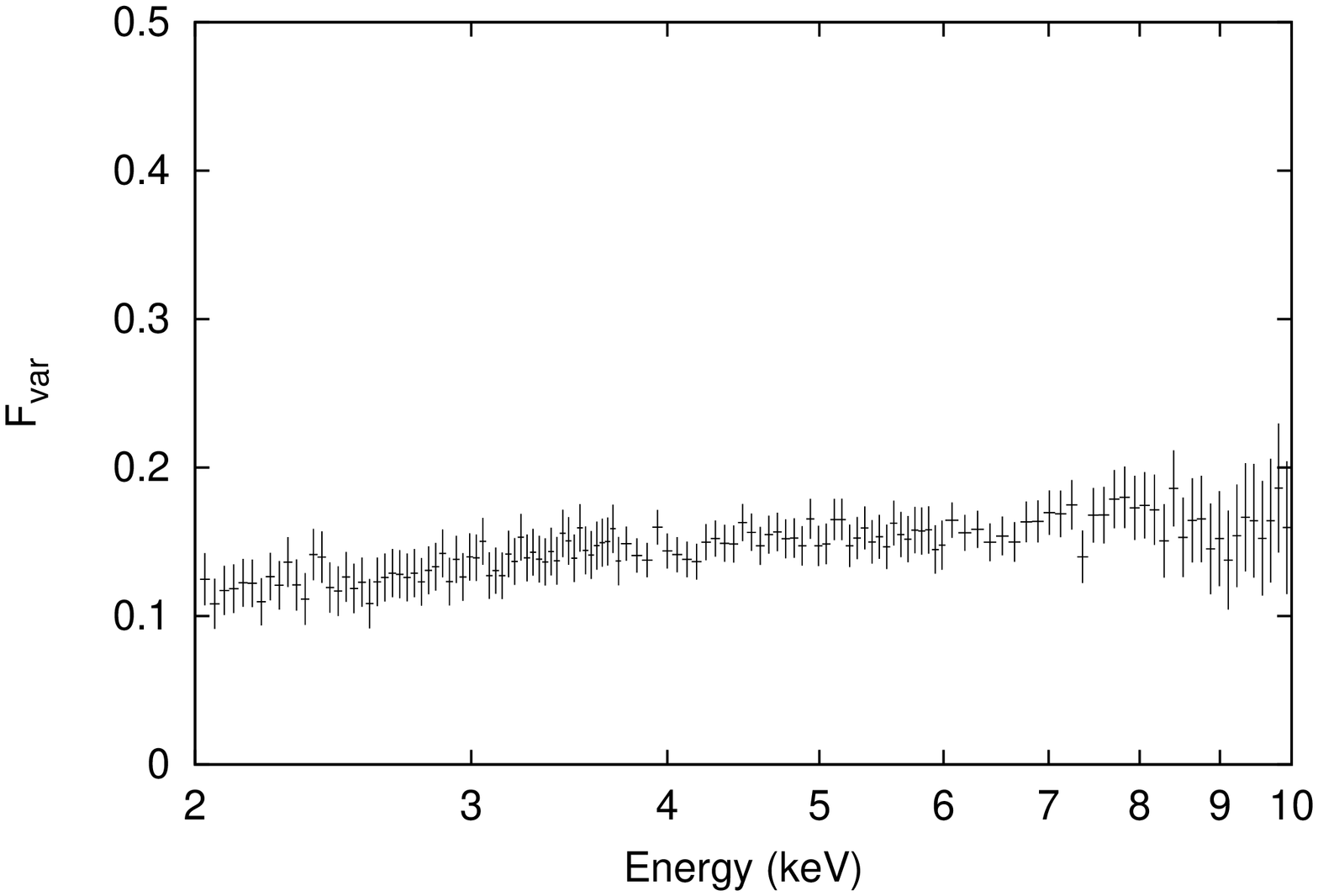}
\caption{Spectral variations of GRS 1915$+$105. The vertical axis shows the variation amplitude with $\Delta T=0.08$~s.}
\label{GRS_DVF}
\end{figure}

Figure \ref{GRS1915} shows the X-ray spectrum of GRS 1915$+$105, and Figure \ref{GRS_DVF} shows the variation amplitude at a timescale of 0.08~s.
We can see {\it no} structure around the iron K-feature,
which is totally different from the variability of AGN (Figure \ref{1H0707_DVF}).
The BH mass of GRS 1915$+$105 is lower than that of 1H0707--495 by 5 orders of magnitude.
If the timescale of spectral variation is normalized by the BH mass,
spectral variations of GRS 1915$+$105 around the iron K-feature at a timescale of 0.08~s would have similar structure to the spectral variations of 1H0707--495, which varies at a timescale of 8000s.
Thus, the observational result is unexpected.
Moreover, we calculate the variation amplitude at any timescales from 60~ms to 63000~s, and {\it no} structures around the iron K-energy band are seen.
This clearly shows that spectral variations around the iron K-energy band are {\it not} normalized by the BH mass.

\section{Discussion}
We have found that the spectral variations of BHs (AGN and BHB) are {\it not} normalized by their masses.
In the relativistic disc reflection model, most variation of the iron feature is caused by change of the source height.
The geometrical configuration should be the same for AGN and BHB if normalized by the Schwarzschild radius, and its physical size would be scaled with the BH mass.
Hence, BHB should show similar variability to AGNs in a few ms \citep{min04}.
Thus, the original relativistic disc reflection model may not explain the observation, 
unless physical conditions of the corona and/or the accretion disc are variable at another timescale by some unknown reasons.

In the partial covering model, 
spectral variability is mostly explained by change of the covering fraction, so the timescale of variability is primarily determined by motion of the partial covering absorbers (e.g.\,\citealt{miy12}).
If the location of partial covering absorbers is more distant from the central BH, the timescale of spectral variations would be longer, and vice versa.
The location of the absorbers is calculated with the following equations:
\begin{equation}
\xi=\frac{L}{nr^2}
\end{equation}
thus
\begin{equation}
r=\left(\frac{L}{n\xi}\right)^{1/2},
\end{equation}
where $\xi$ is ionization parameter, $L$ is the intrinsic X-ray luminosity, $n$ is number density, and $r$ is distance between the central BH and the partial absorbers.
When the size of absorbers along the line-of-sight ($d$) is similar to $r$,
\begin{equation}
r\simeq\frac{L}{\xi N_H},
\end{equation}
where $N_H$ is column density of the absorbers ($N_H=nd$).
In this way, we can estimate $r$ from observables.
In 1H0707--495, $r$ is estimated as $\sim500\,r_s$, while in GRS 1915$+$105, $r\sim10^5\,r_s$ \citep{miz14,miz15}.
This shows that locations of the partial absorbers are not normalized by the BH mass.
Consequently, spectral variations are not normalized by the BH mass, either.

\begin{figure}[tbp]
\centering
\includegraphics[width=55mm,angle=270]{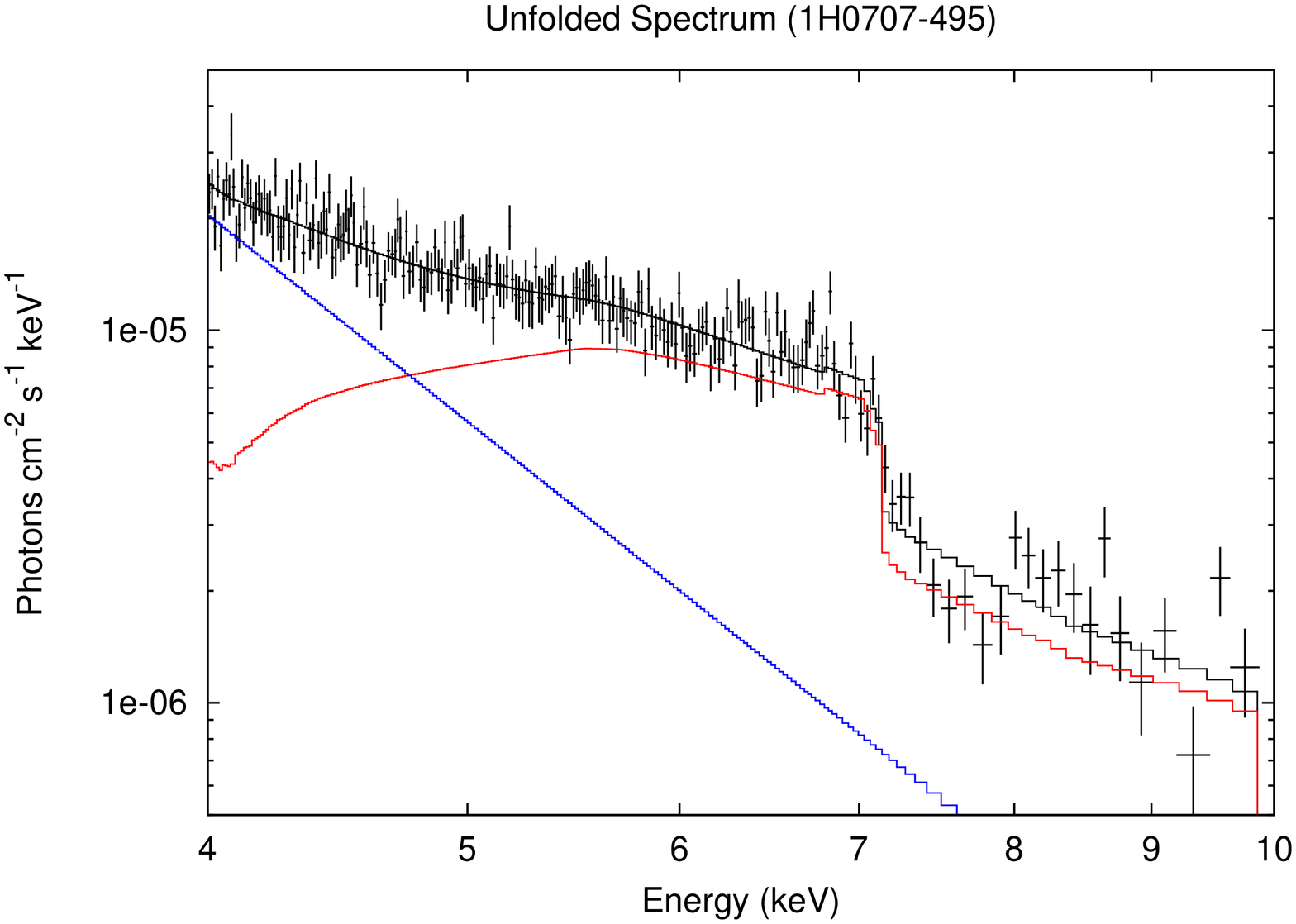}\\
\includegraphics[width=55mm,angle=270]{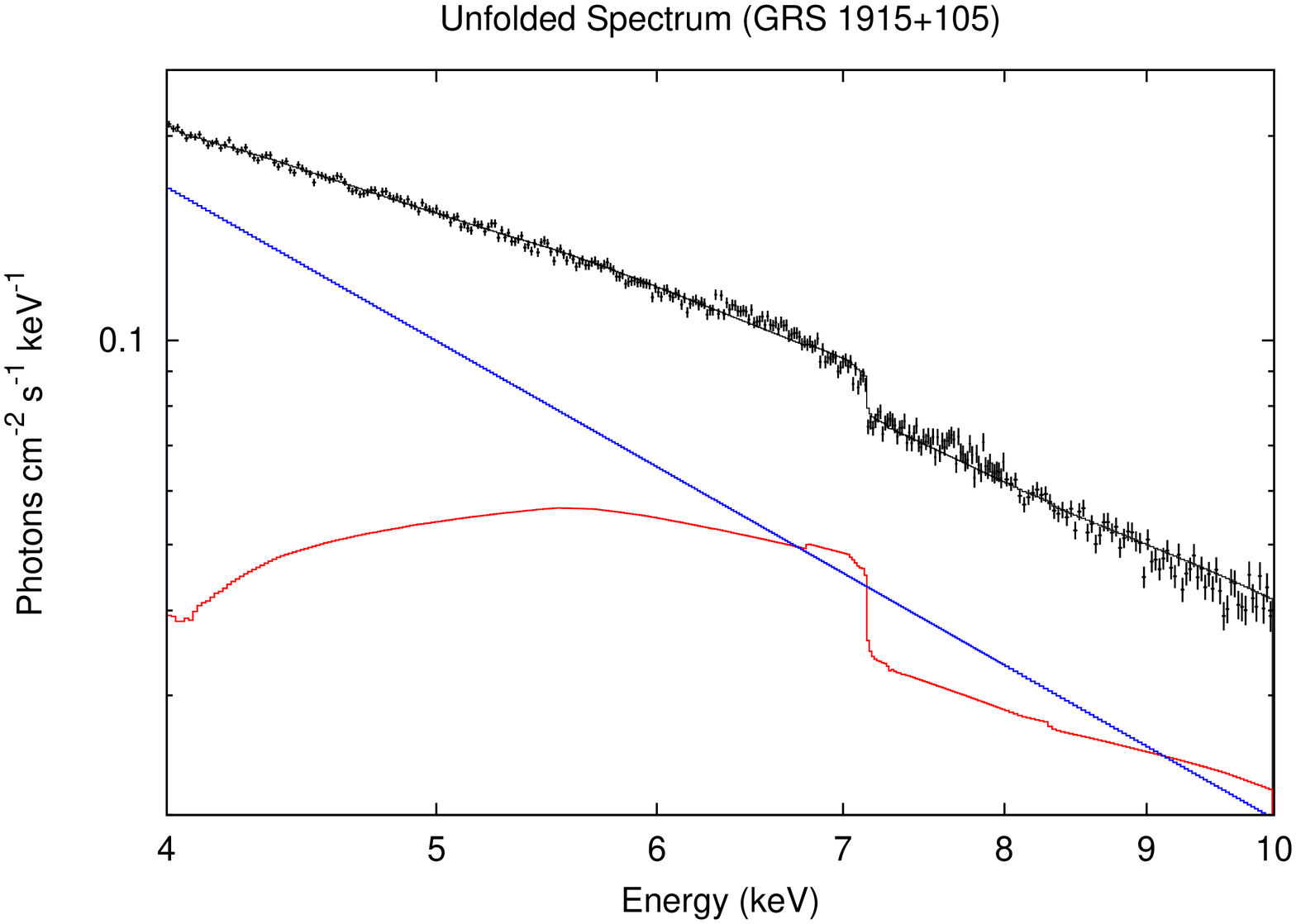}
\caption{Spectral fitting of the broad iron K-spectral features of 1H0707--495 (upper) and GRS 1915$+$105 (lower). The intrinsic X-ray spectra are powerlaw components. The blue lines show unabsorbed components and the red lines show absorbed components.}
\label{nh}
\end{figure}

\begin{table*}
\centering
\caption{Fitting parameters of partial absorbers with the partial covering model.}
\label{tlab}
\begin{threeparttable}
\begin{tabular}{ccc}\hline
 &1H0707--495&GRS 1915$+$105\\ 
\hline
Covering factor &0.96$_{-0.02}^{+0.01}$ & 0.62$\pm$0.02\\
Column density & 1.3$_{-0.2}^{+0.1}\times10^{24}$~cm$^{-2}$& 4.4$\pm$0.5 $\times10^{23}$~cm$^{-2}$\\
Ionization parameter ($\log\xi$) & 1.4$_{-0.3}^{+0.5}$ & 1.09$\pm$0.03\\
\hline
\end{tabular}
\begin{tablenotes}\footnotesize
\item[$\ast$] Errors are quoted at the statistical 90\% level.
\end{tablenotes}
\end{threeparttable}
\end{table*}

Theoretical studies on outflow also suggest that locations of the partial covering absorbers are not normalized by the BH mass between AGN and BHB.
In order to produce the observed iron K-edge feature, the absorbers need not to be fully ionized, and the ionization degree needs to be about $10^1-10^2$ (Figure \ref{nh} and Table \ref{tlab}).
In AGN, a ``radiation-driven'' outflow is considered to take place \citep{ste90,nom13}.
In this outflow, the gas in such a lower-ionization state that can produce the iron K-edge is accelerated by ultraviolet radiation from the disc, thus the wind is launched at the middle of the disc, typically, $\sim400\,r_s$ \citep{nom13}.
On the contrary, in BHB, the accretion disc is hotter than AGN, so absorbers are fully-ionized around the middle of the disc.
Therefore, the ``radiation-driven'' outflow does not work efficiently, and
the ``thermally-driven'' outflow is considered to be dominant \citep{pro02}.
In the ``thermally-driven'' outflow, gas is accelerated by the thermal energy, thus the wind is launched where the binding energy is weaker than the thermal energy \citep{beg83}.
Therefore, the location of the absorbing outflow gas is distant from the central BH.
In this way, the different locations of the partial covering absorbers between AGN and BHB are considered to reflect difference of the outflow mechanisms between them.

\acknowledgements   
The authors would like to thank the Suzaku XIS team for their calibration of P-sum mode.
This work has made use of public Suzaku data obtained through the Data ARchives and Transmission System (DARTS) provided by ISAS/JAXA, and
public XMM-Newton data obtained through the High Energy Astrophysics Science Archive Research Center (HEASARC) at NASA/GSFC.
M.\,M. is financially supported by the Japan Society for the Promotion of Science (JSPS) KAKENHI Grant Number 15J07567.
M.\,T. is financially supported by MEXT/JSPS KAKENHI Grant Numbers 24105007 and 15H03642.



%


\bibliographystyle{aa}
\bibliography{00}

\end{document}